\documentclass[a4paper,showpacs,amsmath,amssymb,12pt]{article}
\usepackage{mathrsfs}
\usepackage{amsmath}
\usepackage{amsfonts}
\usepackage{indentfirst}
\usepackage{amssymb}
\usepackage{latexsym}
\usepackage{graphicx}
\usepackage[english]{babel}
\usepackage{epsfig}
\usepackage{slashbox}
\input epsf.sty

\title{Production of four-quark states with double heavy quarks at LHC
}
\author{Yu-Qi Chen and Su-Zhi Wu\footnote{zzswu@itp.ac.cn} \\
{\small Key Laboratory of Frontiers in Theoretical
Physics,}\\
{\small Institute of Theoretical Physics, Chinese Academy of Sciences} \\
{\small Beijing 100190, P.R. China}}
\date{}

\begin{document}
\maketitle
\begin{abstract}
We study the hadronic production of four-quark states with double
heavy quarks and double light antiquarks at LHC. The production
mechanism is that a color anti-triplet diquark cluster consisting of
double heavy quarks  is formed first from the produced double heavy
quark-antiquark pairs via $gg$ fusion hard process, followed by the
fragmentation of the diquark cluster into a four-quark  (tetraquark)
state. Predictions for the production cross sections and their
differential distributions are presented. Our results show that it
is quite promising to discover these tetraquark states  in  LHC
experiments both for large number events and for their unique
signatures in detectors.

\end{abstract}

\begin{flushleft}
keyword: tetraquark, diquark, heavy quark, fragmentation, production
\end{flushleft}

\section{Introduction}
 LHC experiments provide the unique opportunity to explore some exotic
states of heavy quarks. Among them, particularly interesting states
are those four-quark states consisting of double heavy quarks and
double light antiquarks (or their charge conjugator). The existence
of such states can be inferred from the heavy quark symmetry. The
double heavy quarks in the color anti-triplet state may form a
diquark cluster by the attractive strong interactions. In the heavy
quark limit, the double heavy quarks move slowly in a small relative
velocity $v$ in the rest frame within a smaller distance ($1/mv$),
comparing to the size of light degree of freedom
($1/\Lambda_{QCD}$). Thus in the tetraquark states the double heavy
quarks form a color anti-triplet diquark cluster. It contributes the
color interactions to double light antiquarks as a color source of a
heavy antiquark. The other two light antiquarks move around it with
attractive interactions between them. The dynamics of the light
degrees of freedom of these tetraquark states are similar with those
of the heavy baryons. The picture is supported by scrutinizing the
typical sizes of real hadrons.  The size of the heavy quarkonium is
around $0.2-0.3$ fm while that of the light hadrons is around $1$
fm. The masses of such states can be roughly estimated as the sum of
two heavy quarks' masses and $\Lambda_{QCD}$. For the tetraquark
states containing the $cc$, $cb$, and $bb$ quarks,  their masses are
around 3.4 GeV, 6.8 GeV, and 10.2 GeV, respectively. The flavor
features of this sort of hadrons are very different from the
conventional hadrons. Once they are discovered in experiments, it
will be an undoubted evidence for the existence of the tetraquark
states. Some theoretical studies on these states were presented in
literatures\cite{x2,x3,x4,x5,x6,x7}.

The spin of a tetraquark state is a composition of the spins of the
four quarks and the relative orbital angular momenta between them.
For $S-$waves,  all orbital angular momenta vanish. Then the spin of
the tetraquark state is just the composition of the spin of each
quark or antiquark. The composition of the spins of double heavy
quarks may be 0 or 1. However, when two heavy quarks are identical,
only spin 1 state is allowed due to the antisymmetry by exchanging
identical fermions. It is also the same case for the light
antiquarks sector. In this paper, we are interested in only the
tetraquark states with all orbital angular momenta vanishing. We
denote states by $T_{Q_1Q_2}^i$  ($i$=0, 1) , where $Q_1$ and $Q_2$
represent their heavy flavor indexes  and  $i$ is the spin  of the
double heavy quarks subsystem.

As bound states with quite large masses they are difficult to be
produced at usual high energy machines. Nevertheless, at LHC, they
can be produced efficiently via a direct production process which
involves effects happened at several distinct distance scales.
Firstly, two heavy quark-antiquark pairs are produced via
gluon-gluon fusion  hard subprocess at distance scale $1/m$ or
shorter. Secondly, for those produced two heavy quarks with small
relative velocity $v$ there are certain probabilities to form a
color anti-triplet diquark cluster at distance scale $1/mv$. Finally
the diquark cluster evolves into the tetraquark state  via the
fragmentation process by picking up two light antiquarks from the
vacuum at the distance scale $1/\Lambda_{QCD}$.

In this paper, we calculate the hadronic production cross sections
of $T_{cc}^1$, $T_{bc}^i$ ($i$=0, 1), and  $T_{bb}^1$ via
gluon-gluon fusion process at LHC. Our results show that a number of
these particles can be produced. We also point out the signatures to
detect those particles.

The rest of the paper are organized as follows. In Sec. \ref{sec2},
we present the calculation of the  subprocess $gg \to T^i_{Q_1Q_2}
\bar{Q_1}\bar{Q_2} +X $. Sec. \ref{sec3} devotes to the numerical
results of the cross sections of the tetraquark states at LHC and at
Tevatron. In Sec. \ref{sec4}, we give some discussions about the
results and the signals in the detectors.

\section{Cross section of $gg \to T_{Q_1Q_2}^i
\bar{Q_1}\bar{Q_2} +X $ }\label{sec2}

As mentioned above, in the production process of the tetraquark
states  $T^i_{Q_1Q_2}$, there are three hierarchy  distance scales,
i.e., $ 1/m \ll 1/mv \ll 1/\Lambda_{QCD} $. Accordingly, the cross
sections of the subprocesses for the production of the $S-$wave
tetraquark states, $T_{Q_1Q_2}^{i}$, can be factored into three
different parts accounting for physical effects happening at those
distinct distance scales:
\begin{eqnarray} \label{cs-sub}
\widehat{\sigma}(gg \to T_{Q_1Q_2}^i)
 \,=\, \frac{1}{2\hat{s}}\frac{1}{64d}
 \int d\Pi_{3} \,
 C_{\bar{3}}(\alpha_s,Q_1Q_2) \,
|\Psi_{\bar{3}}(0)|^2 \,
 \int_{0}^{1}dx D _{\bar{3}\to T_{Q_1Q_2}^{i} }(x),
\end{eqnarray}
where $d=1$ for $T_{bc}^{0}$ and $T_{bc}^1$, and $d=2$ for
$T_{bb}^1$ and $T_{cc}^1$; $\hat{s}$ is the squared invariant mass
of the double gluons; $d\Pi_{3}$ is the Lorentz invariant three-body
phase space integral element; $ C_{\bar{3}}(\alpha_s,Q_1Q_2) $ is
the short-distance coefficient describing the production rate of the
color anti-triplet point-like $Q_1Q_2$ state at the energy scale $m$
or higher; $\Psi_{\bar{3}}(0)$ is the wave function at the origin of
the $S-$wave diquark state; $D_{\bar{3}\to T_{Q_1Q_2}^{i} }(x)$ is
the fragmentation function of the diquark into the color-singlet
tetraquark state $T_{Q_1Q_2}^{i}$.

The short-distance coefficient can be calculated by perturbative QCD
and can be expanded in terms of $\alpha_s$ at the short-distance
energy scale $m$ or higher. The rest parts are non-perturbative
effects in nature. To estimate the production cross sections, one
needs to determine their numerical values. The formation of the
diquark cluster from the free double heavy quarks is described by
the wave function at the origin. Their numerical values can be
estimated by the  potential model. The diquark cluster provides an
anti-triplet color source as a heavy antiquark. Thus, in the heavy
quark limit, its fragmentation probability  for forming the
tetraquark states can then be approximately described by that for
forming the heavy baryons by a heavy quark.

In this paper, we calculate the cross sections of hadronic
production of the $S-$wave states, $T_{cc}^1$, $T_{bc}^1$,
$T_{bc}^0$, and $T_{bb}^1$, at LHC. We compute the tree level
short-distance coefficients $C_{\bar{3}}(\alpha_s,Q_1Q_2)$ in the
leading order $\alpha_s^4$ using perturbation QCD. By estimating the
nonperturbative matrix elements, we carry out the numerical
calculations of the total cross sections. Our results show that a
number of these particles can be produced.

We first calculate the short-distance coefficients $
C_{\bar{3}}(\alpha_s,Q_1Q_2) $ at the tree level. They are
proportional to the squared matrix elements $M(gg \to
(Q_1Q_2)_{\bar{3}})$. To calculate the matrix elements, one needs to
calculate the subprocess of $gg\rightarrow{Q_1}{Q_2}
\bar{Q}_1\bar{Q}_2$, with ${Q_1}$ and ${Q_2}$ moving in the same
$3-$velocity and in the color anti-triplet state. At the tree level,
the production processes involve  36 Feynman diagrams for
$gg\rightarrow{b}c\bar{b}\bar{c}$, and 72 ones for
$gg\rightarrow{b}b\bar{b}\bar{b}$ and
$gg\rightarrow{c}c\bar{c}\bar{c}$. The amplitudes  can be classified
as six  gauge-invariant subsets in terms of six independent color
bases. Given this, the calculations of the amplitudes as well as
their squares are straightforward.

$\Psi_{\bar{3}}(0)$'s are the  wave functions at the origin of the
S-wave diquark clusters. Their precise values are difficult to be
gained  since the large range interaction potential between the
double heavy quarks in the color anti-triplet state is not very
clear while the short range one is dominated be  Coulomb potential.
Reasonably, we take values predicted by solving Sch{\"{o}}dinger
equation with Coulomb potential, $v(r)=-{2\alpha_s}/({3r})$. Then
the predicted $\Psi_{\bar{3}}(0)^2$'s are 0.143, 0.0382, and 0.0198
GeV$^3$ for $bb$, $bc$, and $cc$ diquark systems, respectively.
Including the confinement part, the wave function is squeezed to the
central region and hence the wave function at the origin will be
enhanced. Actually, we have done numerical calculations for the
diquark state by solving the Schr{\"{o}}dinger equation using
Coulomb potential plus the linear potential fixed in the
color-singlet case, the numerical value of wave function at the
origin is enhanced by 10\% - 30\%. More reliable prediction for this
nonperturbative number can be obtained by some nonperturbative
method like lattice QCD.

We now turn to  the fragmentation  function of
$(Q_{1}Q_{2})_{\bar{3}}$-cluster to a tetraquark state. As discussed
above, the produced heavy diquark cluster in the color anti-triplet
state provides the same color source as the heavy antiquark  to the
light antiquarks in the limit of the ratio of  size of the diquark
over that of the light antiquarks in the tetraquark state. Thus the
fragmentation probabilities to produce the tetraquark states
$T_{Q_1Q_2}^i$ from the heavy diquarks are the same with that  to
produce the heavy baryons from the heavy quarks. Let's take a QED
example to illustrate it. Imagine the hydrogen ion or deuterium ion
passing material. The probabilities forming the hydrogen atom or the
deuterium one are the same if the velocities of both ions are the
same since they possess the same electric charge. The fragmentation
function is defined in the framework of the infinity momentum where
the parton moves in the speed of light.

The fragmentation  functions to produce the tetraquark states are
nonperturbative in nature. Thus the shapes of them can only be
described by certain phenomenological models \cite{x8,x10}. One of
the most commonly used models is the Peterson model\cite{x8}, in
which the fragmentation function takes the following form:
\begin{eqnarray}\label{frag-sub}
D_{\bar{3}}\to
{T_{Q_1Q_2}^{i}}(x)=\frac{N}{x[1-(1/x)-\epsilon_{Q_1Q_2}/(1-x)]^2}
\end{eqnarray}
where $\epsilon_{Q_1Q_2}$ is  the only parameter determining the
shape of the fragmentation function; $N$ is the normalization
constant. Once the fragmentation probability to produce the
tetraquark state  R is given, $N$ is fixed by the following
condition:
\begin{eqnarray}\label{nor-sub}
 \int
dx \,D_{\bar{3}}\to {T_{Q_1Q_2}^{i}}(x)  =R,
\end{eqnarray}

The fragmentation probabilities  of $c\rightarrow \Lambda_{c}$ and
 $b\rightarrow \Lambda_{b}$ have been measured in $e^+e^-$
collisions \cite{ pdg2006,x11,x12}. According to PDG 2006 \cite{
pdg2006}, $R(c\to\Lambda_c)=0.094\pm0.035$, and
$R(b\to\Lambda_b)=0.099\pm0.017$. These results are about 0.1.
Therefore as a good approximation we may take the fragmentation
probability of $(Q_{1}Q_{2})_{\bar{3}}\rightarrow
T_{Q_{1}Q_{2}}^{i}$, the value of $R$ in Eq.(\ref{nor-sub}), be 0.1.

From Eqs.(\ref{frag-sub}) and (\ref{nor-sub}) we know the
normalization constant $N$ in the Peterson model is dependent on the
parameter $\epsilon_Q$. The model suggested a scaling behavior for
the parameter $\epsilon_Q$ that is proportional to $1/m^2_{Q}$. The
 $\epsilon_{b}$ determined by experiments is about $0.003\sim0.006$
\cite{x11,x13}. Using the scaling behavior and taking $\epsilon_{b}$
to be 0.004, we  predict that
$\epsilon_{bc}=(\frac{m_{b}}{m_{bc}})^2\epsilon_{b}\simeq0.0023$,
$\epsilon_{cc}=(\frac{m_{b}}{m_{cc}})^2\epsilon_{b}\simeq0.011$,
$\epsilon_{bb}=(\frac{m_{b}}{m_{bb}})^2\epsilon_{b}=0.001$ and the
corresponding normalization constants are 0.0075, 0.0194, 0.0047.
Here we take $m_{b}=4.9$ GeV, $m_{c}=1.5$ GeV and
$m_{(Q_{1}Q_{2})}=m_{Q_{1}}+m_{Q_{2}}$.

The fragmentation functions are energy scale dependent. Their
convolution from one energy scale to another satisfies DGLAP
equation. However, for the hadronic production of the tetraquark
states at LHC, the  production is dominated by the smaller $P_T$
region of the produced particles, i.e., compatible with the
tetraquark masses. For smaller $P_T$ production of the tetraquark
state, there are no distinguishable energy scale differences. Thus
the evolution of the fragmentation function used here is not
significant. Therefore, in the numerical calculations we can take
the fragmentation functions with the initial energy scale to be the
tetraquark mass without evolution.

\section{ Numerical results of the cross sections
of the tetraquark states at LHC}\label{sec3}

\begin{figure}
\centering
\includegraphics[width=10cm]{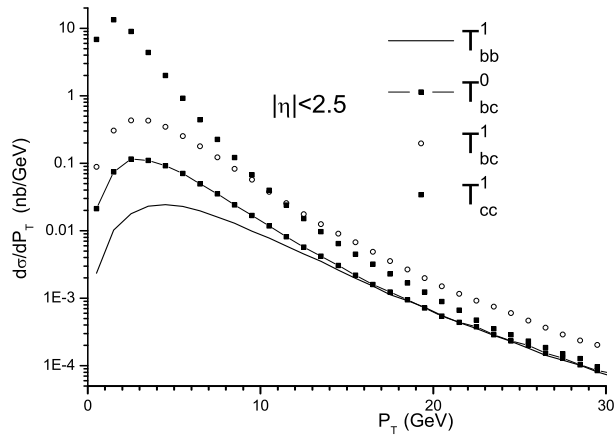}
\caption{The predicted $P_{T}-$distributions of the tetraquark
states, $T_{bb}^1$, $T_{cc}^1$, $T_{bc}^0$ and $T_{bc}^1$, in CMS
and ATLAS with $\sqrt{s}$=14 TeV, with the pseudo-rapidity cut
$|\eta|< 2.5$  and $\mu_F$=$\mu_R$/2.} \label{cmscut}
\end{figure}

\begin{figure}
\centering
\includegraphics[width=10cm]{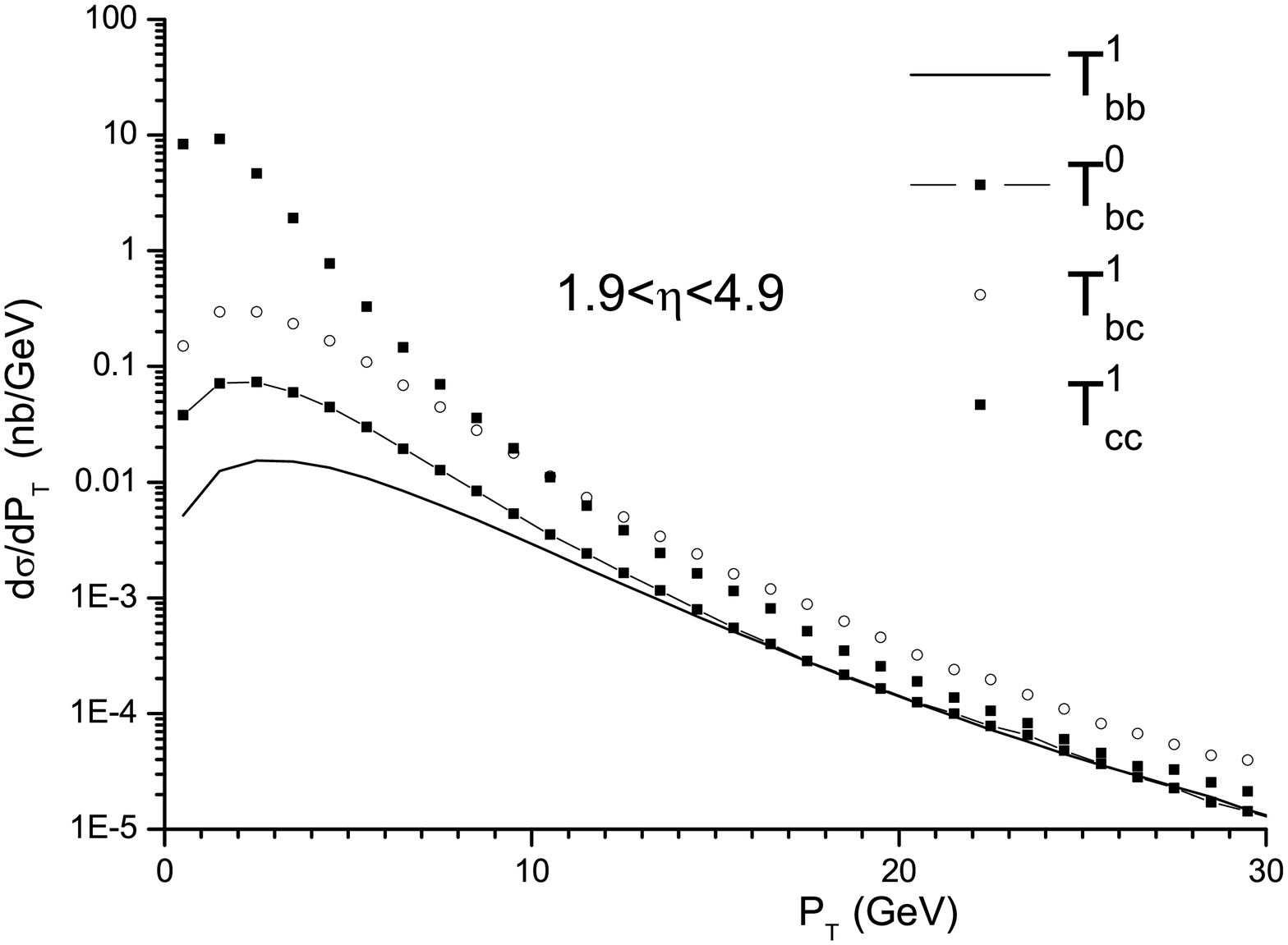}
\caption{The predicted $P_{T}-$distributions of the tetraquark
states, $T_{bb}^1$, $T_{cc}^1$, $T_{bc}^0$ and $T_{bc}^1$, in LHCb
at $\sqrt{s}$=14 TeV, with the pseudo-rapidity cut $1.9<\eta< 4.9$
and $\mu_F$=$\mu_R$/2.} \label{lhccut}
\end{figure}

\begin{figure}
\centering
\includegraphics[width=10cm]{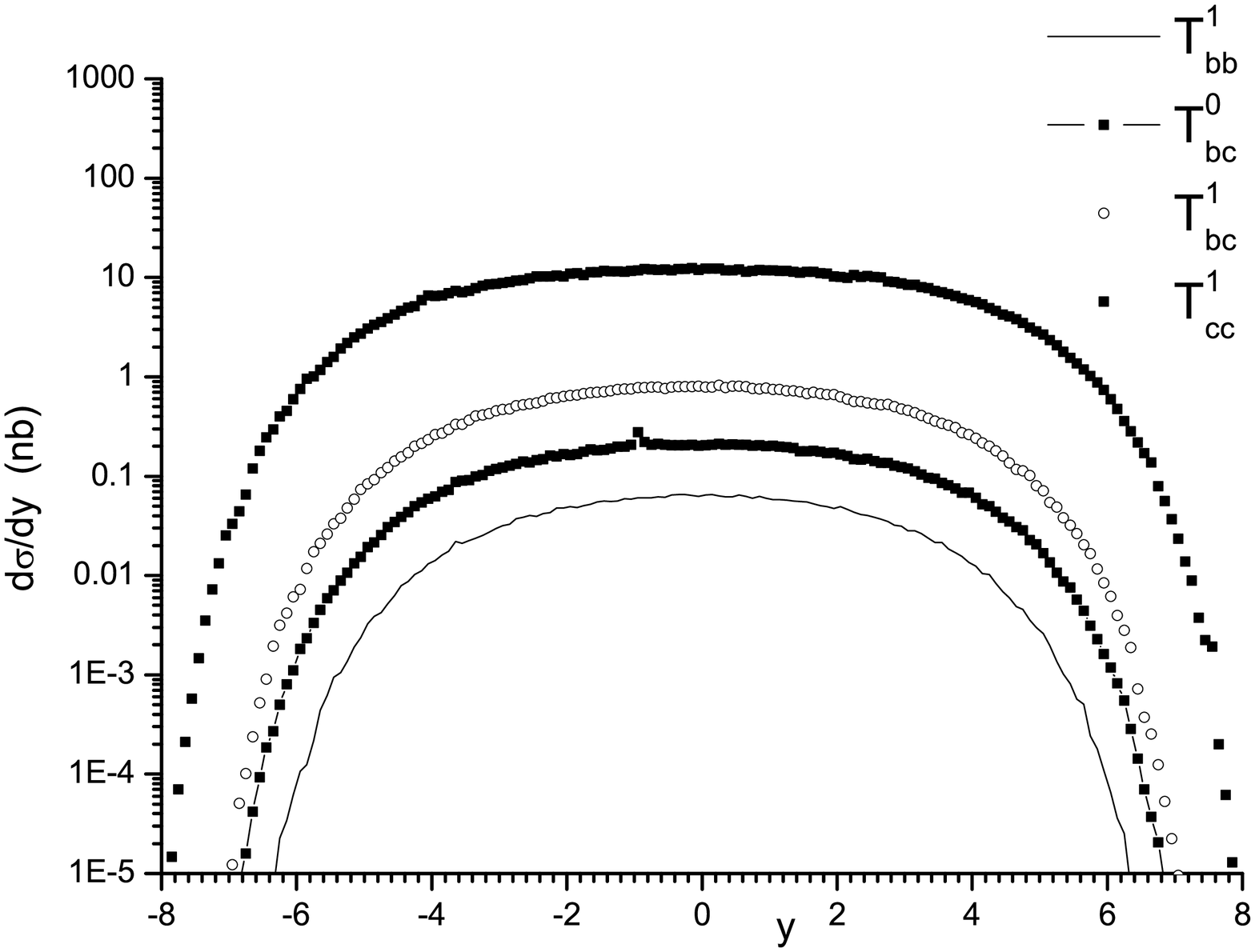}
\caption{The predicted rapidity distributions of the tetraquark
states, $T_{bb}^1$, $T_{cc}^1$, $T_{bc}^0$ and $T_{bc}^1$, with
$\mu_F$=$\mu_R$/2, at LHC with $\sqrt{s}$=14 TeV.} \label{ycut}
\end{figure}

\begin{figure}
\centering
\includegraphics[width=10cm]{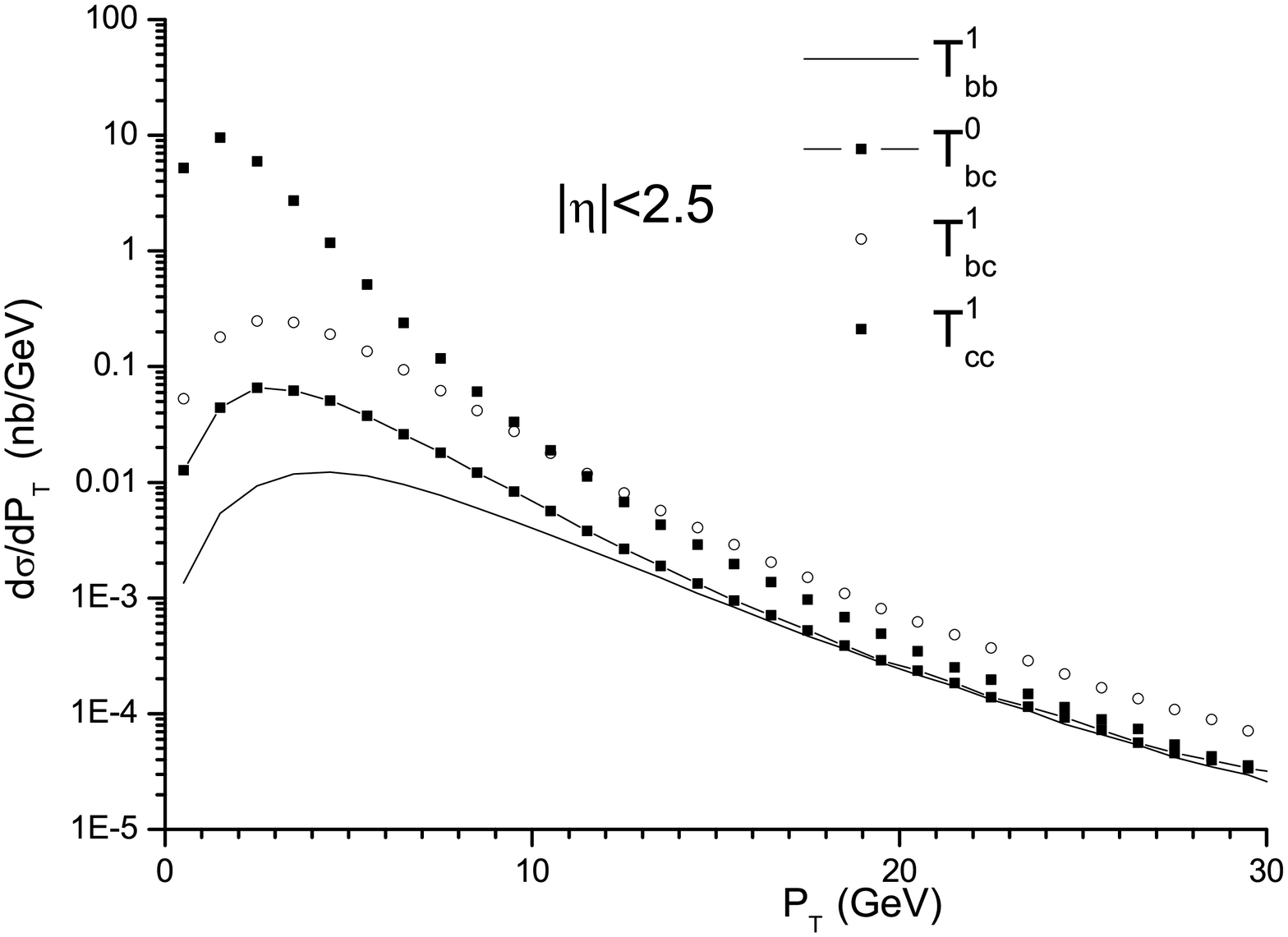}
\caption{The predicted $P_{T}-$distributions of the tetraquark
states, $T_{bb}^1$, $T_{cc}^1$, $T_{bc}^0$ and $T_{bc}^1$, in CMS
and ATLAS at $\sqrt{s}$=7 TeV, with the pseudo-rapidity cut $|\eta|<
2.5$  and $\mu_F$=$\mu_R$/2.} \label{7cmscut}
\end{figure}

\begin{figure}
\centering
\includegraphics[width=10cm]{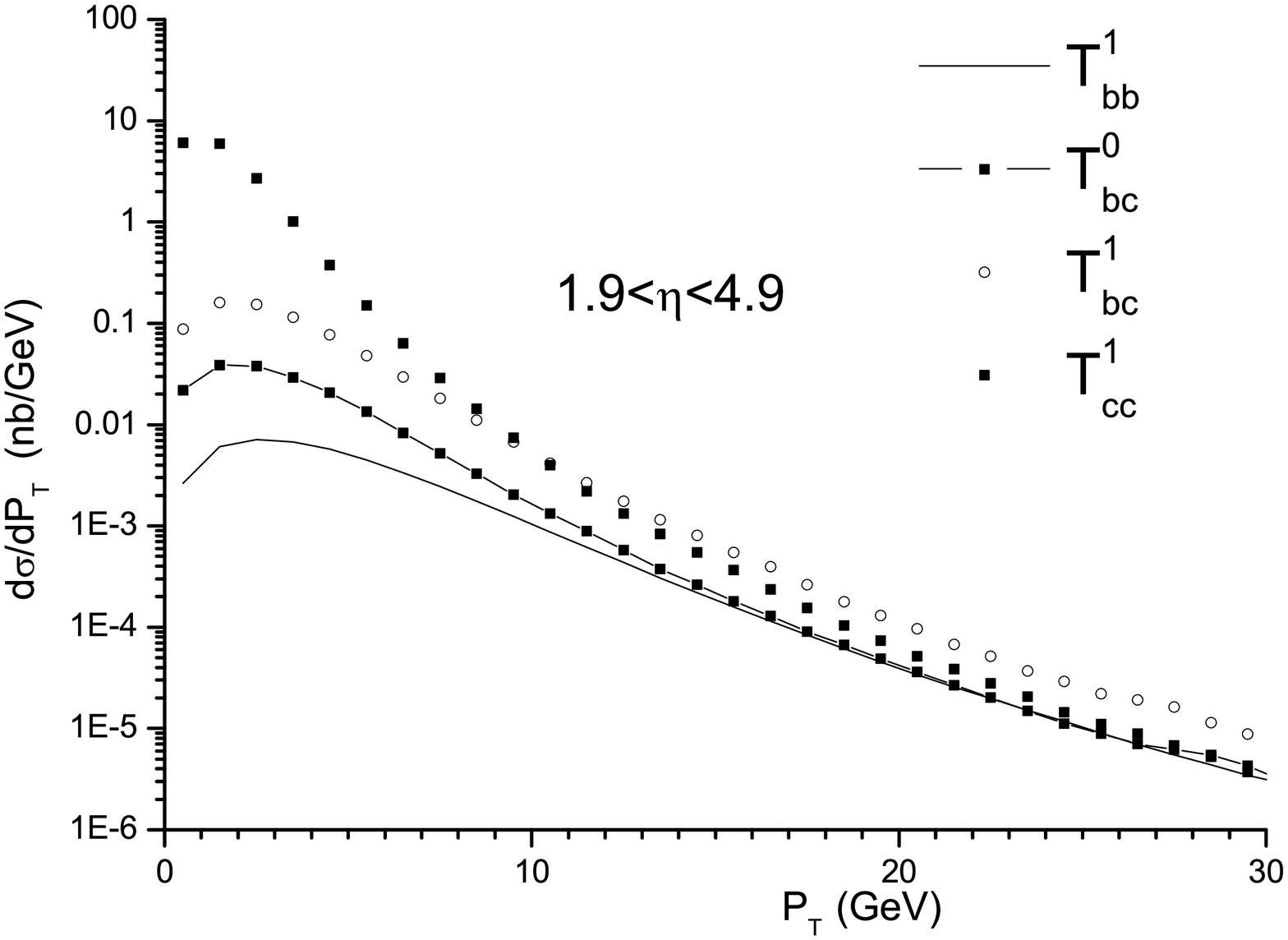}
\caption{The predicted $P_{T}-$distributions of the tetraquark
states, $T_{bb}^1$, $T_{cc}^1$, $T_{bc}^0$ and $T_{bc}^1$, in LHCb
at $\sqrt{s}$=7 TeV, with the pseudo-rapidity cut $1.9<\eta< 4.9$
and $\mu_F$=$\mu_R$/2.} \label{7lhccut}
\end{figure}

\begin{figure}
\centering
\includegraphics[width=10cm]{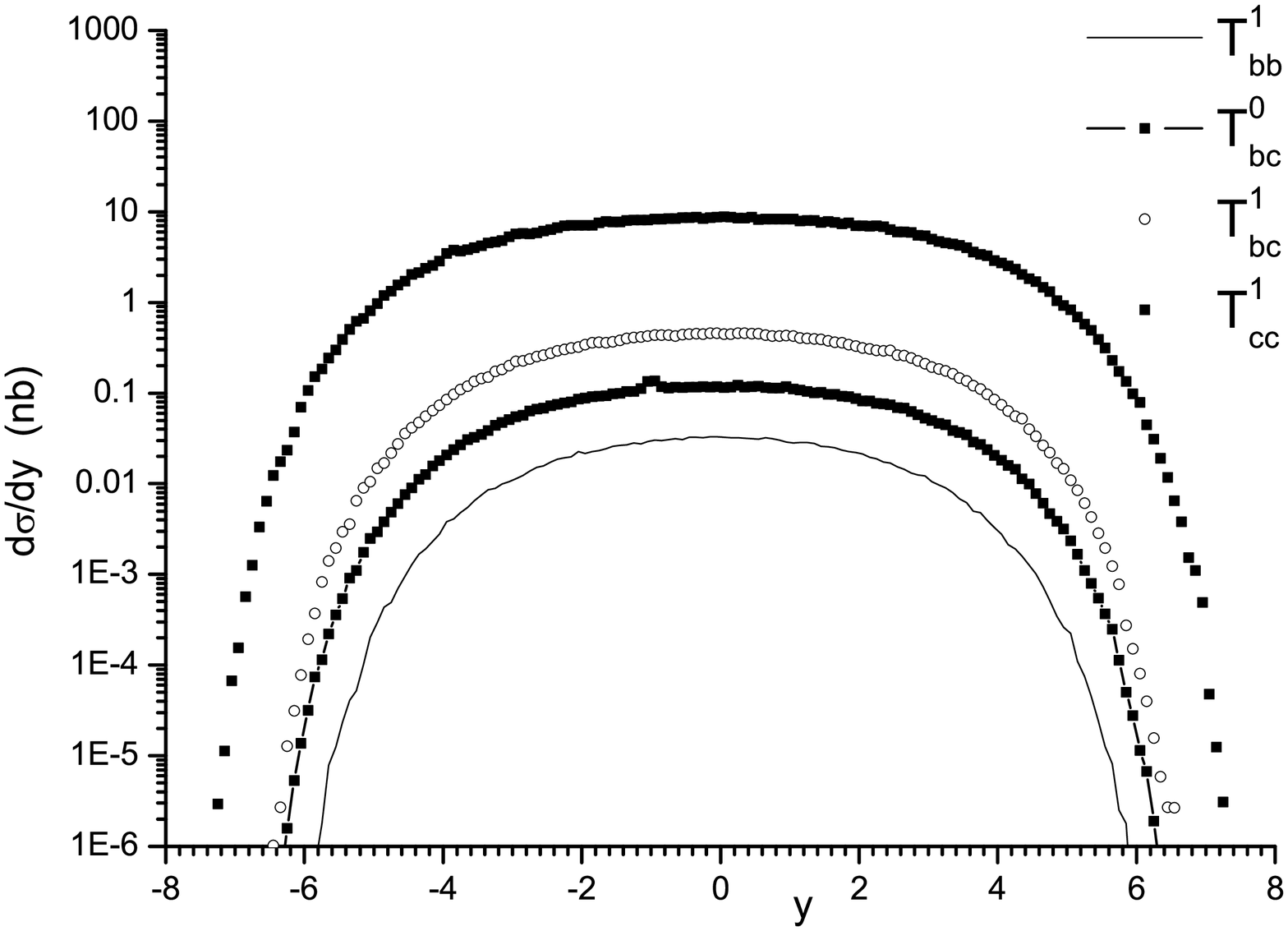}
\caption{The predicted rapidity distributions of the tetraquark
states, $T_{bb}^1$, $T_{cc}^1$, $T_{bc}^0$ and $T_{bc}^1$ with
$\mu_F$=$\mu_R$/2, at LHC with $\sqrt{s}$=7 TeV.} \label{7ycut}
\end{figure}

We now turn to the  calculation of the total cross sections with
$\hat{\sigma}$ of the subprocess given in Eq. (\ref{cs-sub}).
According to the parton model, the cross sections of the processes
$pp\rightarrow T_{Q_{1}Q_{2}}^{i}+{\bar Q_1} +{\bar Q_2}+X$, can be
expressed as:

\begin{eqnarray} \label{total-sub}
\frac{d\sigma}{dP_T}=\int dx_1
dx_2f_{g_{1}}(x_{1},\mu_F)f_{g_{2}}(x_{2},\mu_F)
 \frac{d\widehat{\sigma}}{dP_T}(gg \to T_{Q_1Q_2}^i,\mu_F)\,
 ,
\end{eqnarray}
where $\mu_F$ is the factorization energy scale, $P_T$ is the
transverse momentum of the $T_{Q_1Q_2}^i$, and $f_{g}(x,\mu_F)$ is
the distribution function of the gluon in the proton. Here we use
the cteq6l \cite{x14} parton distribution function. There is an
uncertainty in the calculations arising from the choice of the
factorization energy scale $\mu_F$. Here  for comparison, we take
two different values of $\mu_F$, i.e., $\mu_F=\mu_{R}$ and
$\mu_{R}$/2,  where $\mu_R^2= p^{2}_{TQ_1Q_2}+m^{2}_{Q_{1}Q_{2}}$
with $p_{TQ_1Q_2}$ being the transverse momentum of the diquark.
With the choices of these factorization energy scales and $\alpha_s$
given in \cite{x15} for the cteq6l \cite{x14} parton distribution
function, we calculate the total cross sections in CMS, ATLAS and
LHCb with the invariant mass of the $pp$ system, $\sqrt{s}=7$ TeV
and 14 TeV, respectively. The numerical results of the total
production cross sections are listed in Tables \ref{cross14tev} and
\ref{cross7tev}. The $P_T$ and the rapidity distributions  are shown
in Figs.\ref{cmscut}-\ref{7ycut}. From Tables \ref{cross14tev} and
\ref{cross7tev}, we see  that the total cross section to produce
$T_{cc}^1$ without $P_T-$cut is much larger than the one with
$P_T>5.0$ GeV. This can be explained with two reasons, i.e.,  the
small peak value in the  $P_T$ distribution of the subprocess $gg\to
T_{cc}^1$ and the shape of fragmentation function of $cc-$cluster.
They also result in the lines of the $P_T-$distributions across in
Figs.(\ref{cmscut}),
(\ref{lhccut}), (\ref{7cmscut}), and (\ref{7lhccut}) . 

\begin{table*} \caption{The predicted hadronic production cross
sections (in unit $nb$) of the tetraquark states with various $P_T$
cuts at LHC with $\sqrt{s}$=14 TeV. The pseudo-rapidity cuts
$|\eta|< 2.5$ for CMS and ATLAS, and $1.9<\eta< 4.9$ for LHCb are
taken.}
\begin{tabular}{|c|c|c|c|c|c|}
\hline\hline
-&-& \multicolumn{2}{|c|}{~~LHC (CMS, ATLAS)~~}& \multicolumn{2}{|c|}{LHCb} \\
\hline -&\backslashbox{$P_{T}$-cut} {$\eta$-cut}& \multicolumn{2}{|c|}{$|\eta|< 2.5$} &  \multicolumn{2}{|c|}{$1.9<\eta< 4.9$} \\
\hline \multicolumn{2}{|r|}{$\mu_F$=} & $\mu_R$&$\mu_R/2$& $\mu_R$&$\mu_R/2$\\
\hline $T_{bc}^0$&$0$ GeV&0.460 & 0.643& 0.263&0.374\\
- &$5$ GeV &0.167 &0.238 &0.0590 &0.0865 \\
- &$10$ GeV &0.0286 &0.0429& 0.0077&0.0118 \\
\hline $T_{bc}^1$&$0$ GeV&1.73&2.42&1.02&1.45\\
  -&$5$ GeV &0.567 &0.822& 0.205&0.302 \\
  -&$10$ GeV&0.087 &0.131&0.0235 &0.0356 \\
\hline  $T_{bb}^1$&$0$ GeV&0.129&0.193&0.0697&0.105\\
 - &$5$ GeV& 0.0758&0.114&0.0279 &0.0430 \\
 - &$10$ GeV & 0.0220&0.0339&0.00598 &0.0093 \\
\hline  $T_{cc}^1$&$0$ GeV&35.6&37.8&24.6&25.7\\
 - &$5$ GeV&1.29 &1.84&0.414 &0.629 \\
 - &$10$ GeV&0.072 &0.113& 0.0178&0.0274 \\
\hline\hline
\end{tabular}\label{cross14tev}
\end{table*}

\begin{table*}
\caption{The predicted hadronic production cross sections (in unit
$nb$) of the tetraquark states with various $P_T$ cuts  at LHC with
$\sqrt{s}$=7 TeV . The pseudo-rapidity cuts $|\eta|< 2.5$ for CMS
and ATLAS, and $1.9<\eta< 4.9$ for LHCb are taken. }
\begin{tabular}{|c|c|c|c|c|c|}
\hline\hline
-&-& \multicolumn{2}{|c|}{~~LHC (CMS, ATLAS)~~}& \multicolumn{2}{|c|}{LHCb} \\
\hline -&\backslashbox{$P_{T}$-cut} {$\eta$-cut}& \multicolumn{2}{|c|}{$|\eta|< 2.5$} &  \multicolumn{2}{|c|}{$1.9<\eta< 4.9$} \\
\hline \multicolumn{2}{|r|}{$\mu_F$=} & $\mu_R$&$\mu_R/2$& $\mu_R$&$\mu_R/2$\\
\hline $T_{bc}^0$&$0$ GeV& 0.228&0.352&0.117&0.184\\
- &$5$ GeV &0.0766 & 0.121&0.0221 & 0.0358\\
- &$10$ GeV &0.0119 &0.0195& 0.00243& 0.00407\\
\hline $T_{bc}^1$&$0$ GeV&0.860&1.33&0.456&0.717\\
  -&$5$ GeV &0.265 &0.416&0.0776&0.125 \\
  -&$10$ GeV&0.0364&0.0599&0.0075 & 0.0126\\
\hline  $T_{bb}^1$&$0$ GeV&0.0578&0.0936&0.0273&0.0448\\
 - &$5$ GeV&0.0328 &0.0535&0.0098 &0.0164\\
 - &$10$ GeV&0.0088&0.0145&0.00180 & 0.00304\\
\hline  $T_{cc}^1$&$0$ GeV&20.8&25.9&13.3&16.3\\
 - &$5$ GeV&0.630 &1.01&0.166 & 0.272\\
 - &$10$ GeV &0.030 &0.051& 0.00580& 0.0099\\
\hline\hline
\end{tabular}\label{cross7tev}
\end{table*}

From Table \ref{cross14tev}, we see that for LHC with $300\;
fb^{-1}$ integrated luminosity running at $14$ TeV, for the
production of $T^1_{cc}$, around $(3.9-5.5)\times 10^{8}$ events in
CMS and ATLAS can be accumulated with kinematic cuts $P_T > 5$ GeV
and $|\eta| < 2.5$, while this number is around $(1.2-1.9)\times
10^{8}$ with kinematic cuts $P_T > 5$ GeV and $1.9< \eta < 4.9$ in
LHCb. For the production of $T_{bc}$, both $T^0_{bc}$ and $T^1_{bc}$
need to add together since the higher mass state will decay into the
lower mass state by emitting a photon. After doing this, we see that
for $300\; fb^{-1}$ integrated luminosity running at $14$ TeV, for
the production of $T_{bc}$, around $(2.2-3.2)\times 10^{8}$ events
in CMS and ATLAS can be accumulated with kinematic cuts $P_T > 5$
GeV and $|\eta| < 2.5$, while this number is around $(0.8-1.2)\times
10^{8}$ with kinematic cuts $P_T > 5$ GeV and $1.9< \eta < 4.9 $ in
LHCb. For the production of $T^1_{bb}$, around $(2.2-3.4)\times
10^{7}$ events in CMS and ATLAS can be accumulated with kinematic
cuts $P_T > 5$ GeV and $|\eta| < 2.5$, while this number is around
$(0.8-1.3)\times 10^{7}$ with kinematic cuts $P_T > 5$ GeV and $1.9<
\eta < 4.9 $ in LHCb.

From Table \ref{cross7tev}, we see that for $10\; fb^{-1}$
integrated luminosity running at $7$ TeV, for the production of
$T^1_{cc}$, around $(0.6-1.0)\times 10^{7}$ events in CMS and ATLAS
can be accumulated with kinematic cuts $P_T > 5$ GeV and $|\eta| <
2.5$, while this number is around $(1.6-2.7)\times 10^{6}$ with
kinematic cuts $P_T > 5$ GeV and $1.9< \eta < 4.9 $ in LHCb.  For
the production of $T_{bc}$, both $T^0_{bc}$ and $T^1_{bc}$ need to
add together also. After doing this, we see that for $10\; fb^{-1}$
integrated luminosity running at $7$ TeV, for the production of
$T_{bc}$, around $(3.4-5.4)\times 10^{6}$ events in CMS and ATLAS
can be accumulated with kinematic cuts $P_T > 5$ GeV and $|\eta| <
2.5$, while this number is around $(1.0-1.6)\times 10^{6}$ with
kinematic cuts $P_T > 5$ GeV and $1.9< \eta < 4.9 $  in LHCb. For
the production of $T^1_{bb}$, around $(3.3-5.4)\times 10^{5}$ events
in CMS and ATLAS can be accumulated with kinematic cuts $P_T > 5$
GeV and $|\eta| < 2.5$, while this number is around $(1.0-1.6)\times
10^{5}$ with kinematic cuts $P_T > 5$ GeV and $1.9< \eta < 4.9 $ in
LHCb.

\begin{figure}
\centering
\includegraphics[width=10cm]{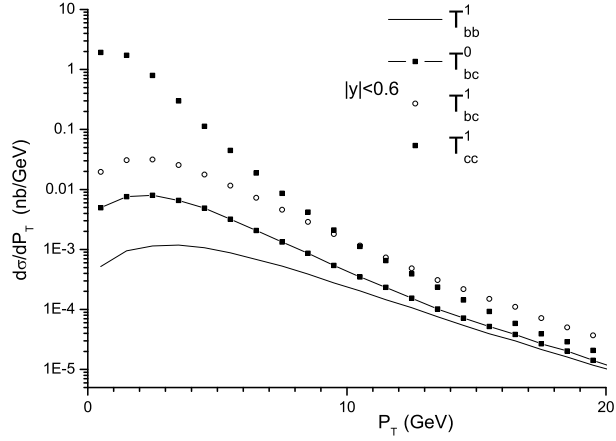}
\caption{The predicted $P_{T}-$distributions of the tetraquark
states, $T_{bb}^1$, $T_{cc}^1$, $T_{bc}^0$ and $T_{bc}^1$, at
Tevatron with $\sqrt{s}$=1.96 TeV, with the rapidity cut $|y|<0.6$
and $\mu_F$=$\mu_R$/2.} \label{tevapt}
\end{figure}
\begin{figure}
\centering
\includegraphics[width=10cm]{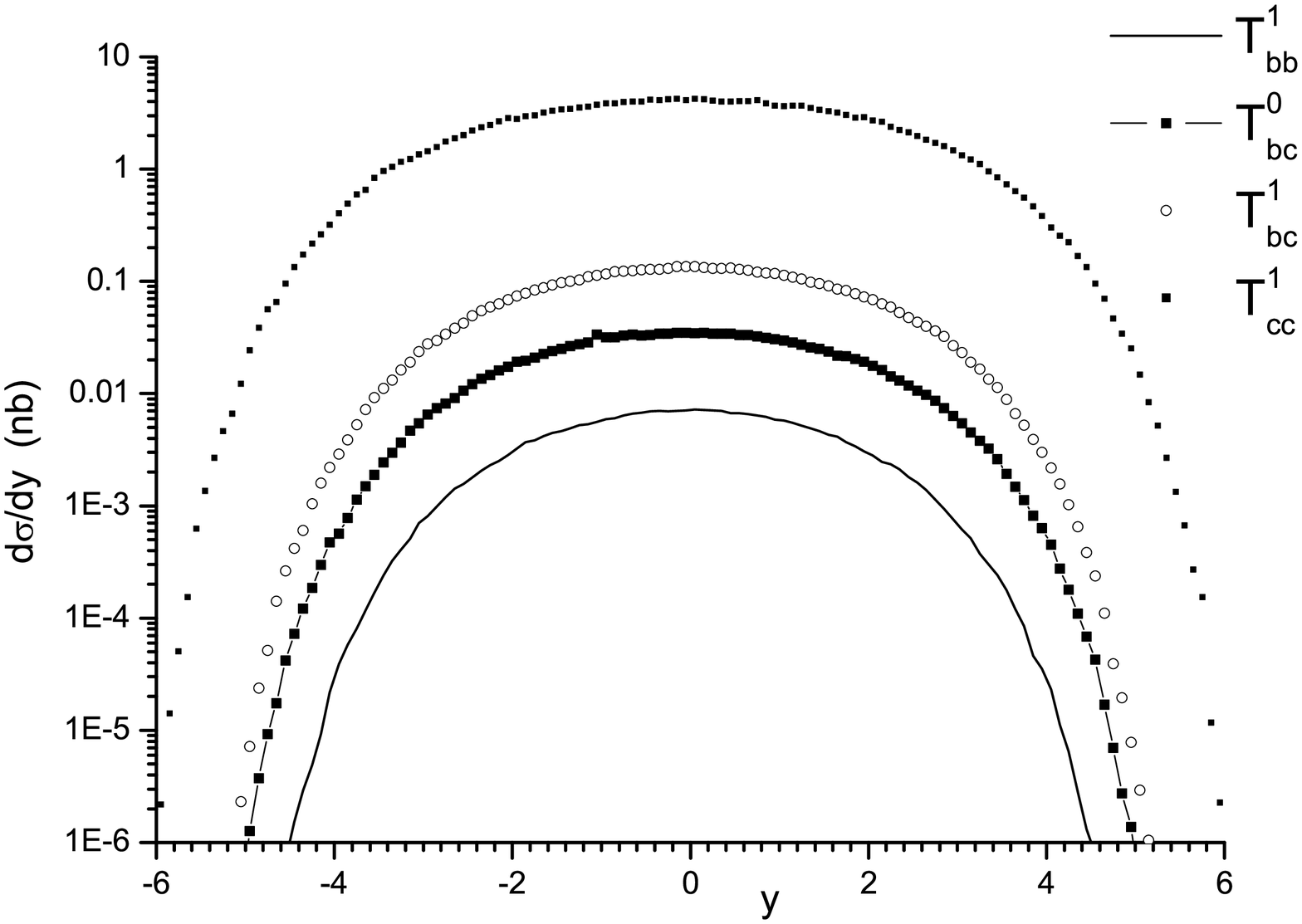}
\caption{The predicted rapidity distributions of the tetraquark
states, $T_{bb}^1$, $T_{cc}^1$, $T_{bc}^0$ and $T_{bc}^1$ with
$\mu_F$=$\mu_R$/2, at Tevatron with $\sqrt{s}$=1.96 TeV.}
\label{tevaeta}
\end{figure}

We also calculate the total cross sections at Tevatron with
$\sqrt{s}=1.96$ TeV . The numerical results of the total production
cross sections are listed in Table \ref{tevatron}. The
$P_T-$distributions and the rapidity distributions  are shown in
Figs.\ref{tevapt} and \ref{tevaeta}.

\begin{table*}[h]
\caption{The predicted hadronic production cross sections (in unit
$nb$) of the tetraquark states with various $P_T$ cuts at Tevatron
with $\sqrt{s}$=1.96 TeV .The rapidity cut $|y|< 0.6$. }
\begin{tabular}{|c|c|c|c|c|c|}
\hline\hline
$\mu_F$&$P_T$ cut&$T_{cc}^1$&$T_{bc}^1$&$T_{bc}^0$&$T_{bb}^1$ \\
\hline $\mu_R$& 0 GeV&3.01&0.0832&0.0219&0.00436\\
-&$5$ GeV &0.041&0.0168&0.00487&0.00182\\
-&$10$ GeV &0.00149&0.00178&0.00058&0.00039\\
\hline $\mu_R$/2& 0 GeV&5.01&0.157&0.0410&0.00837\\
-&$5$ GeV &0.081&0.0312&0.0090&0.00352\\
-&$10$ GeV &0.0029&0.00339&0.0011&0.00075\\
\hline\hline
\end{tabular}\label{tevatron}
\end{table*}

 From Table \ref{tevatron}, we see that for $2\; fb^{-1}$ integrated
luminosity running at $1.96$ TeV, for the production of $T^1_{cc}$,
around $(0.8-1.6)\times 10^{5}$ events at Tevatron can be
accumulated with kinematic cuts $P_T > 5$ GeV and $|y| < 0.6$.  For
the production of $T_{bc}$, again by adding both $T^0_{bc}$ and
$T^1_{bc}$ events together, we see that for $2\; fb^{-1}$ integrated
luminosity, for the production of $T_{bc}$, around $(0.4-0.8)\times
10^{5}$ events at Tevatron can be accumulated with kinematic cuts
$P_T > 5$ GeV and $|y| < 0.6$. For the production of $T^1_{bb}$,
around $0.4-0.7\times 10^{4}$ events at Tevatron can be accumulated
with kinematic cuts $P_T > 5$ GeV and $|y| < 0.6$.

 There are some uncertainties in the predicted numerical results.
One arises from the ambiguity in choosing the factorization energy
scale $\mu_F$. From the Tables, we see that two different values of
$\mu_F$ lead to around and less than factor 2 difference in the
cross section. The next leading order result will reduce this
uncertainty. However, it will be very hard calculation. Another one
arises from the wave function at the origin, as discussed in Sec.
\ref{sec2}, it will increase the cross sections by 20-70 \%
including the linear confinement potential between the heavy quark.
Moreover, all excited heavy diquark states will decay into the
ground diquark states by emitting photons or $\pi$'s.  Including
those contributions, the production cross sections may be enhanced
by 2-3 times.

Comparing the $P_T$ distributions at LHC and Tevatron, we see that
the differential cross sections decrease faster at Tevatron than at
LHC with the $P_T$ increase.

In calculating the accumulated events number, we  take the
integrated luminosity as the corresponding collider running at
present or planning luminosity for about one year. From the above
discussions, we see that the predicted event number of the
tetraquark at the Tevatron is about 3-4 orders of magnitude lower
than that at LHC with $\sqrt{s}=14$ TeV for one year.

The events can be discovered at Tevatron is much smaller than the
ones can be discovered at LHC, especially when the energy and
luminosity of LHC reach the maximum.

\section{Signature and summary}\label{sec4}
The decays of the tetraquark states with double heavy quarks possess
unique signatures in the detectors. To analyze them in detail is
very useful for probing them in experiments. For the ground
tetraquark states, they can not decay by the strong and the
electro-magnetic interactions. However, they can decay through the
weak interaction. The inclusive decay can be regards as the cascade
weak decays of the heavy quarks while the light antiquarks are
spectators. When the leptons can easily be identified in detectors,
we focus on those semileptonic decays of the heavy quarks in the
tetraquark states.

Now we analyze the semileptonic decays of the $T_{Q_{1}Q_{2}}^{i}$
states one by one. By weak interaction, the $c$ quark can decay to
$s$ quark by emitting the $e^{+}$ or $\mu^{+}$, and an invisible
neutrino. In the $T_{cc}^1$, double $c$ quarks decay in this way
independently, with the two antiquarks as spectators. The emitted
same plus sign double leptons and the  $P_T$-distributions can be
used to deduce the background.

In the $T_{bc}^i$, there are two heavy quarks with different
flavors. Both the $c$ and the $b$ quarks have the semi-leptonic
decay modes. As discussed in the last paragraph, the $c$ quark can
decay to $s$ quark by emitting, $e^{+}$ or $\mu^{+}$, and an
invisible neutrino. The $b$ quark can decay to the $c$ quark by
emitting the $e^-$, $\mu^-$,  or $\tau^-$ lepton and the
corresponding invisible anti-neutrino, followed by the $c$ quark
decays as above. As a result, the two plus sign leptons and one
lepton can be detected from those cascade decays, giving the special
signature for identifying the $T_{bc}^i$.

The inclusive semi-leptonic decay of $T_{bb}^1$ is somewhat more
complicated. Each $b$ quark can  decay to the $c$ quark with a
lepton, $e^-$, $\mu^-$,  or $\tau^-$ and the corresponding invisible
anti-neutrino, followed by the $c$ quark decays as above. These
leptons, two leptons and two plus sign leptons, provide the unique
signature for identifying the $T_{bb}^1$ state.

The decay vertexes and the $b-$tagging method will be very useful
for identity those tetraquark states. Some other useful decays for
identifying those tetraquark states are those 3-body or 4-body
nonleptonic decays.

 Considering the signatures and the efficiencies of the the
detectors, we see that LHC with high luminosity running at 14 TeV
provides best opportunity to discover doubly heavy tetraquark
states. In conclusion, our results show that it is quite promising
to discover these tetraquark states in LHC experiments both for
large number events and for their unique signatures in detectors.
Once they are discovered in experiments, it will be an undoubted
evidence for the existence of the tetraquark states. This will start
a new stage in hadron physics.\\
 {\bf Acknowledgments:} This work is partly
supported  by the NSFC with the contract 10875156.


\end{document}